\documentclass[a4paper,jou,natbib,floatsintext]{apa6}

\usepackage[british]{babel}
\usepackage[utf8]{inputenc}
\usepackage[hidelinks]{hyperref}
\usepackage[document]{ragged2e}

\usepackage{parskip}
\setcitestyle{aysep={},yysep={;}}

\graphicspath{ {./figs/} }

\title{Towards decolonising computational sciences}
\shorttitle{Towards decolonising computational sciences}

\leftheader{Birhane and Guest}

\twoauthors{Abeba Birhane}{Olivia Guest}
\twoaffiliations{School of Computer Science, \\ University College Dublin, Ireland \& \\ Lero, the Science Foundation Ireland Research Centre for Software, Ireland}{Research Centre on Interactive Media,  Smart Systems \\ and Emerging Technologies --- RISE, \\ Nicosia, Cyprus \& \\ Department of Experimental Psychology,\\ University College London, UK}


\abstract{
\justify
This article sets out our perspective on how to begin the journey of decolonising computational fields, such as data and cognitive sciences.
We see this struggle as requiring two basic steps:
$a$) realisation that the present-day system has inherited, and still enacts, hostile, conservative, and oppressive behaviours and principles towards women of colour (WoC);
and $b$) rejection of the idea that centering individual people is a solution to system-level problems.
The longer we ignore these two steps, the more ``our'' academic system maintains its toxic structure, excludes, and harms Black women and other minoritised groups. This also keeps the door open to discredited pseudoscience, like eugenics and physiognomy.
We propose that grappling with our fields' histories and heritage holds the key to avoiding mistakes of the past.
For example, initiatives such as ``diversity boards'' can still be harmful because they superficially appear reformatory but nonetheless center whiteness and maintain the status quo.
Building on the shoulders of many WoC's work, who have been paving the way, we hope to advance the dialogue required to build both a grass-roots and a top-down re-imagining of computational sciences --- including but not limited to psychology, neuroscience, cognitive science, computer science, data science, statistics, machine learning, and artificial intelligence.
We aspire for these fields to progress away from their stagnant, sexist, and racist shared past into carving and maintaining an ecosystem where both a diverse demographics of researchers and scientific ideas that critically challenge the status quo are welcomed. 
}

\begin{document}

\maketitle


\section{Introduction}

\begin{quote}
    The most powerful weapon in the hands of the oppressor is the mind of the oppressed.
    
    \cite{Biko1978}
\end{quote}

\vspace{1em}
\justify

In this article, we tackle two related stumbling blocks for the healthy and safe progression and retention of people of colour (PoC) in general in the computational sciences --- fields including but not limited to machine learning (ML) and artificial intelligence (AI), as well as data and cognitive sciences within the Western context.
We intersectionally shed light on the perspectives and experiences of both cis and/or binary (men and women) as well as queer, trans, and non-binary PoC, and we especially focus on women of colour (WoC) and Black women in the computational sciences \citep[][]{collective1983combahee, Crenshaw1990}.
Firstly, we provide an overview of the conservative and hostile status of these fields to PoC and especially to Black people.
The present scientific ecosystem sustains itself by rewarding work that reinforces its conservative structure.
Anything and anyone seen as challenging the status quo faces systemic rejection, resistance, and exclusion.
Secondly, we explain how centering individual people, as opposed to tackling systemic obstacles, is a myopic modus operandi and indeed part of the way the current hegemony maintains itself.
Fundamental change is only possible by promoting work that dismantles structural inequalities and erodes systemic power asymmetries.

As we shall explain, ``our'' current scientific ecosystem is so potent, pervasive, and forceful that even Black women can become assimilated, or at least project assimilationist viewpoints (i.e., integrating into and upholding the status quo).
As such, the current Western computational sciences ecosystem --- even when under the guise of equity, diversity, and inclusivity --- reinforces behaviours (even in Black women) that can be useless to or even impede the healthy progress of (other) Black people within it \citep{OkunND, Chang_2019}.
Black women, through years of training and enculturation in a white supremacist and colonialist system, are conditioned to internalize the status quo. 
They may thus be unable to describe and elucidate the systems that oppresses them.
Even when Black women are able to reckon with their oppression and marginalisation, as their experience is misaligned with the academic value system, they might lack the language to articulate it.
Furthermore, they might be subject to corrective punishment, or at least coercion, to cease further ``rebellion'' \citep{Agathangelou2002}.

We plan to unpack all the above with an eye towards a collective re-imagining of the computational sciences.
To do this, we implore computational scientists to be aware of their fields' histories \citep{Winston2020, Syed2020, Saini2019, Roberts2020, Cave2020} and we propose that through such an awakening we can begin to forge a decolonised future.
We also hope our article encourages researchers to consciously avoid repeating previous mistakes, some of which are crimes against humanity, like eugenics \citep[][]{Saini2019}.
Ultimately, our goal is to make inroads into radically decolonised computational sciences \citep[cf.][]{Birhane2019, Cave2020}.

\section{The computational sciences ecosystem}

\begin{quote}

What does it mean when the tools of a racist patriarchy are used to examine the fruits of that same patriarchy? It means that only the most narrow parameters of change are possible and allowable.

\cite{Lorde1984}
\end{quote}

Computational and cognitive sciences --- fields that both rely on computational methods to carry out research as well as engage in research of computation itself --- are built on a foundation of racism, sexism, colonialism, Anglo- and Euro-centrism, white supremacy, and all intersections thereof \citep[][]{Lugones2016, Crenshaw1990}.
This is dramatically apparent when one examines the history of fields such as genetics, statistics, psychology, etc., which were historically engaged in refining and enacting eugenics \citep[][]{Winston2020, Syed2020, Saini2019, Roberts2020, Cave2020}.
``Great'' scientists were eugenicists, e.g., Alexander Graham Bell,
Cyril Burt,
Francis Galton,
Ronald Fisher,
Gregory Foster,
Karl Pearson,
Flinders Petrie,
and Marie Stopes
\citep{BernalLlanos2020}.


The Western cis straight white male worldview masquerades as the invisible background that is taken as the ``normal'', ``standard'', or ``universal'' position \citep{ahmed2007phenomenology}.
Those outside it are racialised, gendered, and defined according to their proximity and relation to colonial whiteness \citep{Lugones2016}.
People who are coded as anything other than white, have limited to no access to the field, as reflected in the demographics from undergraduate courses to professorships \citep{Roberts2020, Gabriel2017}.
In other words, the current situation in the computational sciences remains one of de facto \emph{white supremacy}, wherein whiteness is assumed as the standard which in turn allows white people to enjoy structural advantages, like access to (higher paying) jobs and positions of power \citep[][]{Myers2018}. 
Mutatis mutandis for \emph{masculine supremacy}: men enjoy structural benefits and privileges, as reflected in the (binary) gender ratios throughout the computational sciences \citep[][]{Huang2020, Hicks2017, Gabriel2017}.

\begin{figure}
    \centering
    \includegraphics[width=\linewidth]{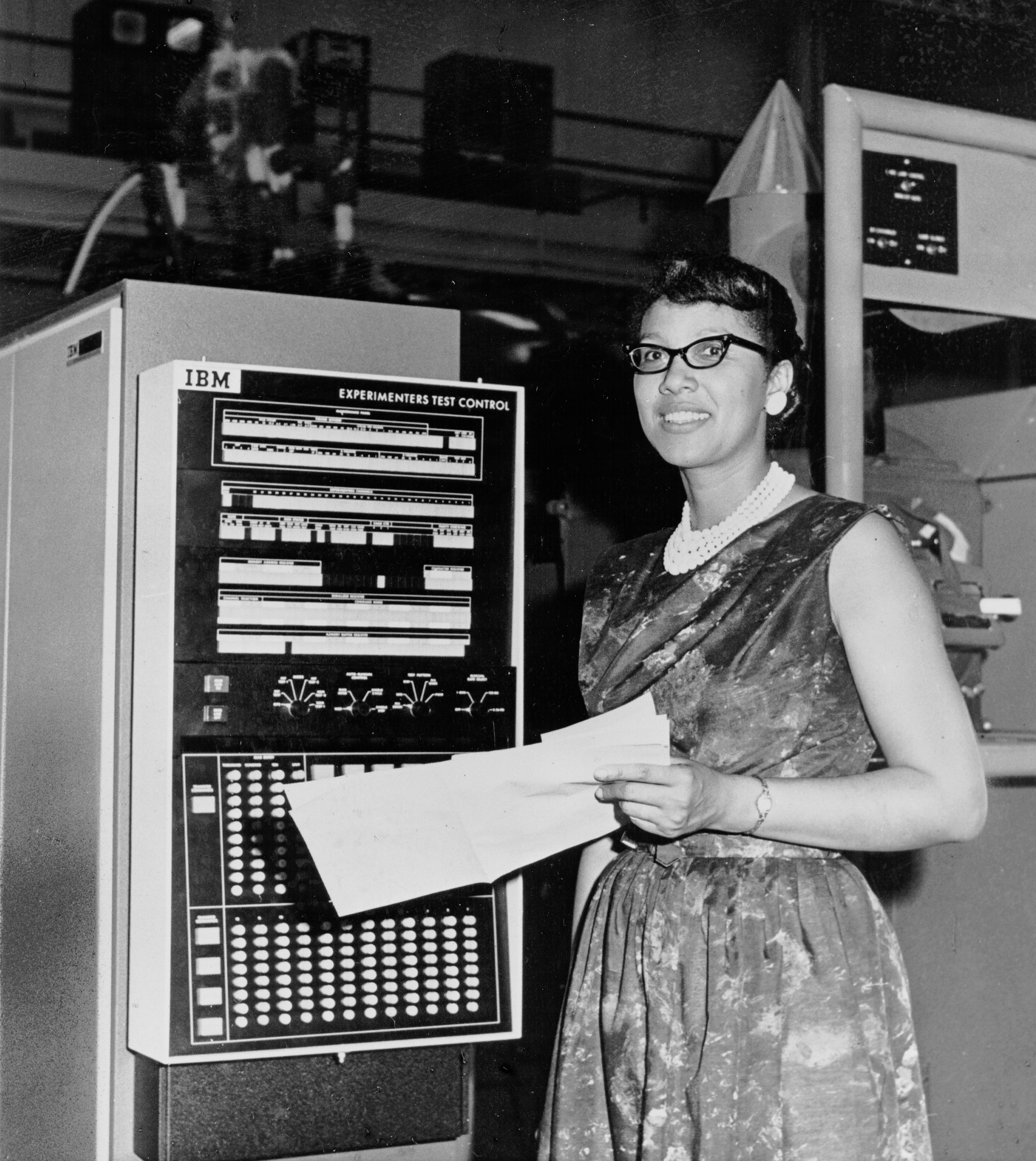}
    
    \vspace{2mm}
    \caption{
    ''Melba Roy Mouton was Assistant Chief of Research Programs at NASA's Trajectory and Geodynamics Division in the 1960s and headed a group of NASA mathematicians called ``computers''. Starting as a mathematician, she was head mathematician for Echo Satellites 1 and 2, and she worked up to being a Head Computer Programmer and then Program Production Section Chief at Goddard Space Flight Center.''
    \citep[photograph by NASA, released to the public domain, ][]{bwic_2016}
    }
    \label{fig:mouton}
\end{figure}

Academia, and science specifically, is seen by some as a bastion of Leftism and so-called ``cultural Marxism'' \citep{Mirrlees2018}, operating to exclude conservativism \citep{ha2020}.
However, both in terms of its demographic make-up and in terms of what are considered ``acceptable'' and ``legitmate'' research endeavours, science is conservative, even within broader Left-leaning ideologies and movements \citep[][]{Mirowski2018}.
This is especially apparent when we consider that many positions of social and political power reflect the broader demographics of the  societies in which scientific institutions are embedded, while these same scientific institutions lag behind in terms of representation.
For example, in terms of political power, 10\% of MPs in the UK are minoritized ethnic, reflecting the 13.8\% of people in the UK with a non-white background \citep{Uberoi2019}.
Similarly, in the USA, 27.2\% of the members of the House of Representatives are minoritized ethnic while 23.5\% of the USA population identifies as such  \citep{Uberoi2019}.
Science's ability to grant positions of power to minorities is abysmal in comparison.
In 2017, there were only 350 Black women professors in the UK across all fields, making up less than 2\% of the professoriate and five out of
159 University Vice Chancellors (3.1 \%) are Black \citep[]{Khan2017, Linton2018}.
Relatedly, Black women's writings are systemically omitted from syllabi and Black women have to work extra hard  --- producing higher levels of scientific novelty --- to get the equivalent recognition and reward to white men \citep{Hofstra2020}.
Historically, Black women, even more than women in general, have been erased making evidence of their pioneering work and leadership within computational sciences, like Melba Roy Mouton (see \autoref{fig:mouton}), difficult to find \citep{Hicks2017, Nelsen2017}.

Due to computational sciences' history --- especially our lack of institutional self-awareness, which protects hegemonic interests --- white and male supremacy continues to sneak (back) into even seemingly sensible research areas.
For example, under the guise of a seemingly scientific endeavour, so-called ``race science'' or ``race realism'' conceals much of the last two centuries' white supremacy, racism, and eugenics \citep{Saini2019}.
Despite a wealth of evidence directly discrediting this racist pseudoscience, race realism  --- the eugenic belief that human races have a biologically based hierarchy in order to support racist claims of racial inferiority or superiority --- is currently experiencing a rebirth, chiefly aided by AI and ML \citep[e.g.,][]{Arcas2017}.
Computational sciences in general, and AI and ML specifically, hardly examine their own histories --- apparent in the widespread ignorance of the legacies of research on IQ and on race studies from the fields of statistics, genetics, and psychology \citep[e.g.,][]{BernalLlanos2020, Cell2020, Laland2020, Winston2020, Syed2020, prabhu2020large, Cave2020}.
As we currently stand, AI and ML are best seen as forces that wield power where it already exists, perpetuating harm and oppression \cite[]{kalluri2020don}. 

In the present, harmful discredited pseudoscientific practices and theories like eugenics, phrenology, and physiognomy, even when explicitly promoted, face little to no push back \citep{Saini2019, chinoy2019racist, stark2018facial}.
Springer, for example, was recently pressured to halt publication of a physiognomist book chapter.
Scholars and activists wrote an extensive rebuttal which was then signed by over two thousand 
experts from a variety of fields \citep{cct2020}.
No official statement was provided condemning such work by the editors or publishers, despite being explicitly called on to condemn this type of pseudoscience. Regardless, Springer Link continues to publish pseudoscience of similar magnitude. At the time of writing, for example, we identified 47 papers published this year (2020) alone by Springer Link, all claiming to have built algorithmic systems that ``predict gender'' even though the very idea of predicting gender has been demonstrated to rest on scientifically fallacious and ethically dubious grounds \cite[]{keyes2018misgendering}. 
This exemplifies how seemingly progressive actions function as fig-leaves obfuscating and preserving the system's conservatism, white supremacy, and racism.

The lack of field-wide, top-down critical engagement results in an uptick in publications that revive explicit scientific racism and sexism \citep[][]{birhane2019algorithmic, prabhu2020large}.
Tellingly, such ideas are defended not via deep ideological engagement or coherent argumentation but by appealing to
rhetorical slights of hand.
In the rare cases where papers are retracted following outrage, it is the result of a large effort often spearheaded by researchers who are junior, precarious, and/or of colour \citep[e.g.,][]{Gliske2020, Mead2020}.
A much higher energy barrier is needed to get such flawed work expunged from the academic record than is needed to slip such work into the literature in the first place.
Unfortunately, the retraction of a few papers, in a publishing culture that fails to see the inherent racist, sexist, and white supremacist, foundations of such work serves only as a band-aid on a bullet wound. 
The system itself needs to be rethought --- scholars should not, as a norm, need to form grassroots initiatives to instigate retractions and clean up the literature. Rather, the onus should fall on those producing, editing, reviewing, and funding (pseudo)scientific work. Strict and clear peer review guidelines, for example, provide a means to filter racist pseudoscience out \cite[]{boyd2020racism}. Ultimately, it is the peer review and publishing system, and the broader academic ecosystem that need to be re-examined and reimagined in a manner that explicitly excludes harmful pseudoscience and subtly repackaged white supremacist from its model. 

In the present, white supremacism, racism, and colonialism are promoted through (increasingly) covert means and without the explicit consent of most research practitioners nor human participants. White supremacist ideological inheritances, for example, are found in subtle forms in modern academic psychological, social, and cognitive sciences \citep[][]{Winston2020, Syed2020, Roberts2020}.
Many of the conclusions about the so-called ``universal'' human nature are based on the observations of people from societies that are described as Western, educated, industrialized, rich, and democratic \citep[WEIRD,][]{Henrich2010}.
Although this appears as an obvious form of white supremacy --- where a select few are deemed representative of the whole human experience --- nonetheless, practitioners have often been oblivious until the default way of collecting data has been described in explicit terms.

In a similar manner, colonialism in academia does not take on the form of physical invasion through brute force \citep[][]{George2002, Birhane2019}.
Instead we are left with the remnants of colonial era mentality:  coloniality \citep{mohamed2020decolonial}.
There is no mainstream direct advocacy for (neo-)Nazi propaganda, for example, but there is facilitation of the CIA's torture programme \citep{Welch2017, Soldz2011}.
Additionally, there are prominent and/or tenured academics who promote anything from support of the status quo to palingenesis \citep[return to an idealised past,][]{Griffin2018}, collectively known as the Intellectual Dark Web (IDW), e.g.,
Jonathan Haidt,
Sam Harris,
Christina Hoff Sommers,
Jordan Peterson, Steven Pinker, and
Bret Weinstein \citep{Ribeiro2020, Parks2020}.
These researchers use their academic credentials to promote conservative to alt-right ideologies to their large public following, including the idea that science is actively hostile to their ideas while subsequently calling for ``civility'' in the face of hate \citep{ha2020}.
According to the IDW, leftism and liberalism are the dominant frameworks in science.
This is a useful rhetorical device for upholding the status quo, akin to a systemic-variant of a tactic called DARVOing: deny, attack, and reverse victim and offender \citep[][]{Harsey2017}.


\section{A tale of two academias}

\begin{quote}
    When confronted with something that does not fit the paradigm we know, we are likely to resist acknowledging the incongruity.
    
    \cite{Onuoha2020}
\end{quote}
\vspace{1em}

Academia's oppressive structures are invisible to those in privileged positions --- the matrix of oppression \citep{Ferber2007} is rendered transparent, undetectable.
This holds, in some cases, even for minoritized scholars who are trained in fields like the computational sciences where oppressive forces and troubling foundations are not the subject of scrutiny. 
Concepts and ideologies set out by a homogeneous group of ``founding fathers'' or ``great men'' are presented as ``objective'', ``neutral'', and ``universal'', seemingly emerging from ``the view from nowhere'' and obscuring the fact that they embody the status quo.
Interrogating the history and underlying assumptions of these concepts is often seen as political and/or ethical and, therefore,
outside the purview of scientific enquiry.
This blocks the attempts of Black women --- whose experience is not captured by so-called universal concepts --- to carve out an academic home.

For those who satisfy, and are satisfied with, the status quo, academia is ``comfortable, like a body that sinks into a chair that has received its shape over time'' \citep{Ahmed2014}.
Noticing how the chair might be uncomfortable for others is a difficult task even when its uncomfortableness has been explicitly demonstrated.
The recent \#BlackInTheIvory hashtag on Twitter \citep[][]{Subbaraman2020a} illustrates how dramatically painful the Black academic experience is (quoted with permission):

    

    
\begin{quote}
    \#BlackintheIvory As faculty member in an institution, guard wouldn't let me in the library. Showed my faculty ID, [with] my photo. ``Is that really you?''
    
    Mario L. Small (@MarioLuisSmall)
\end{quote}
\vspace{1em}

\begin{quote}
    The confusion on your students face, at the start of every semester when you walk into a classroom, with the realization that a black [woman] will be teaching them. \#BlackInTheIvoryTower
    
    Abeba Birhane (@Abebab)
\end{quote}
\vspace{1em}

\begin{quote}
    To white/non-[Black, Indigenous, and PoC] folks in academia asking yourself if you ever contributed to the things being discussed in \#BlackintheIvory, let me assure you that the answer is yes. It was probably just something so inconsequential to you that you don't even remember it.
    
    Naomi Tweyo Nkinsi (@NNkinsi)
\end{quote}
\vspace{1em}

\begin{quote}
   On the rare occasions (before I knew better) that I shared my \#BlackintheIvory experiences [with] colleagues who were not Black, it usually led to invalidation and gaslighting. So to see this out in the open is incredible, but it surfaces pain that I continually suppress to survive.
   
   Jamila Michener (@povertyscholar)
\end{quote}

The \#BlackInTheIvory hashtag demonstrates that despite operating within the general umbrella of ``academia'', Black scholars face radically different treatment compared to their non-Black counterparts --- they inhabit a dramatically more hostile environment.
They are under constant scrutiny, evaluated according to divergent, more stringent, standards \citep{Spikes2020}.
This hostile parallel environment otherises minoritized academics and remains imperceptible, even unimaginable, to privileged academics.

\begin{figure}[bt]

\setlength{\tabcolsep}{2pt}
\renewcommand{\arraystretch}{1}
\centering
\newcommand\x{0.315}

\begin{tabular}{lll}
\textsf{\textbf{a}) ground truth \vspace{1mm}} 
&   \textsf{\textbf{b}) blurred input} 
&  \textsf{\textbf{c}) output} \\

\includegraphics[width=\x\linewidth]{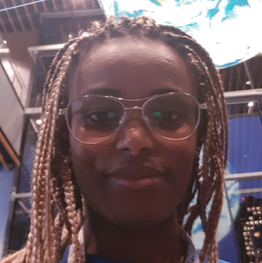}
 &  \includegraphics[width=\x\linewidth]{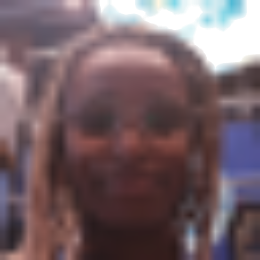}
 &  \includegraphics[width=\x\linewidth]{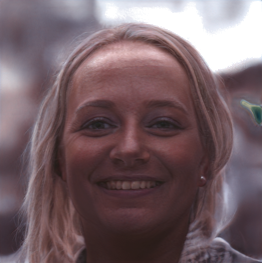} \\

  \includegraphics[width=\x\linewidth]{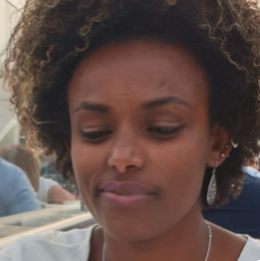}
 &  \includegraphics[width=\x\linewidth]{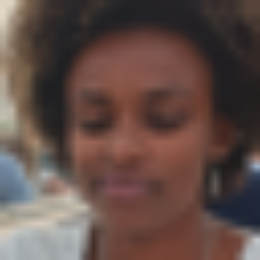}
 &  \includegraphics[width=\x\linewidth]{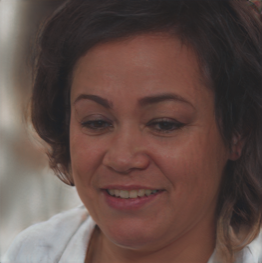}
 \\

  \includegraphics[width=\x\linewidth]{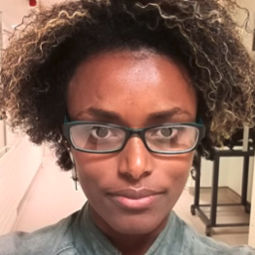}
 &  \includegraphics[width=\x\linewidth]{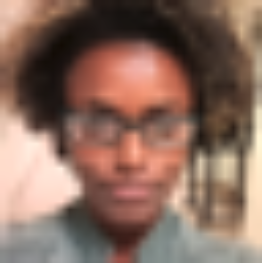}
 &  \includegraphics[width=\x\linewidth]{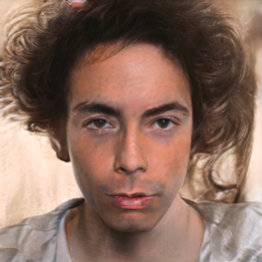}
 \\
\end{tabular}
\caption{Three examples of Abeba Birhane's face (column a) run through a depixeliser \citep{Menon2020}: input is column b and output is column c.
}
\label{fig:whiteface}
\end{figure}

Oftentimes, Black women's attempts to describe their lived reality and their request for fair and just treatment is met with backlash typically from white, cis, male, etc., academics, both in senior and junior positions.
Black women exist under a near constant threat of misogynoir, the intersection of sexism and anti-Blackness \citep{Bailey2018}.
From being labeled ``angry'', ``loud'', and ``nasty'', to being demeaned with phrases such as ``it is a subjective experience, not an objectively verifiable claim'' \citep{Walley_Jean_2009}.
Black women are even more obviously gaslit, i.e., their concerns are discarded systematically, leading to them doubting their reality and judgements of the toxicity of the system \citep{Davis_2017}.
On the one hand, individual cases of racism are dismissed as one-off instances that cannot be evidential for structural racism.
On the other hand, overarching patterns of racism are deemed irrelevant on the basis that specific cases cannot be characterised based on aggregate data.
These two rhetorical devices allow for undermining Black women and for explaining away misogynoir.
When those in positions of power accept anecdotal evidence from those like themselves, but demand endless statistics from minoritized groups, no amount of data will suffice \citep{Lanius2015}.

Computational scientists who are both Black and women face daily mega- to microaggressions involving their intersectional position \citep[][]{Sue2007}. Take this seemingly banal algorithm that depixelises images 
, for example. When confronted with a Black woman's face, it ``corrects'' her Blackness and femininity, see \autoref{fig:whiteface}.
This type of erasure exemplifies the lack of a diverse team, the lack of a diverse testing-stage userbase, and a deep lack of understanding about how imposing digital whiteface constitutes harm, i.e., is a(n micro)aggression \citep[]{Sloane2020}. But more fundamentally, and far from being an isolated incident of lack of proper testing and imagination, 
this is a symptom of the subtle white and male supremacy under which the computational fields operate, which assume and promote whiteness and maleness as the ideal standards.

\section{White women are part of the problem}

\begin{quote}
    White feminism is the feminism that doesn't understand western privilege, or cultural context. It is the feminism that doesn't consider race as a factor in the struggle for equality.
    
    \cite{Young2014}
\end{quote}
\vspace{1em}

Perhaps unsurprisingly, diversity cannot realistically be achieved by merely focusing on gender diversity.
When the existence of oppressive systems is acknowledged within the computational fields, it is common for institutions to assemble ``diversity and inclusion boards'', often composed of white women.
The reasoning behind this typically amounts to ``women are victims of an oppressive academic system, therefore, their active involvement solves this problem''. Such discourse is reflective of the institutional ineptitude at thinking beyond individualised solutions and towards systems-level change.
This oversimplified approach is naive, and even harmful \cite{chang2019mixed}.
The assumption that, cisgender heterosexual ablebodied Western, white women represent all women is misguided \citep[]{ahmed2007phenomenology}.    

White women are beneficiaries of all the advantages that come with whiteness --- white supremacy, coloniality, Orientalism, and Anglo- and Eurocentrism.
White feminism, i.e., feminism that is anti-intersectional, cannot address these issues \citep{Young2014}.
White feminism is a one-size-fits-all ideology that decries centring issues other than (a narrow definition of) patriarchy, claiming that such deviations are divisive.
For example, white feminism is loathe to, and indeed not equipped to, discuss the coloniality of the gender binary \citep{Lugones2016}.
Importantly, although white feminism is mainly advanced by its beneficiaries --- white women --- it it not limited to being enacted purely by white women.
It can be inherited and internalized regardless of racialisation, which means that white feminism has to do more with the ideology than gender, race, or ethnicity \citep[see ]{Nadar2014}.

As we discuss in the previous section, oppressive structures are difficult to see and understand for people who do not occupy a certain racialized and politicized space --- ``where the chair is not made in their mould''. White women are often unable to detect white supremacist, Eurocentric, and colonial systems.
This has implications for progress or rather, it hiders progress. 
The centering of white women, especially those who explicitly advance white feminism, does not remedy structural problems --- no single individual can.
White feminist actors also monopolize, hijack, and even weaponise, these spaces, deflating multi-dimensional and hierarchical intersectional issues, e.g., misogynoir, and reducing them into a single dimension, stripped of all nuance, of the oppressive system they face: the patriarchy \cite[]{eddo2018m}. 
This manifests in defensiveness and hostility, like the use of canned phrases such as ``not all white women'', when Black women point out oppression beyond the patriarchy.
Ultimately, we all need to ask ourselves: ``How can decades of feminist epistemology and more recently Black feminist epistemology and research practice enhance research practice in general and not just the practices of those who self-identify as feminists?'' \citep[][p. 20]{Nadar2014}


\section{Tokenism and its discontents}

\begin{quote}

One way of excluding the majority of Black women from the knowledge-validation process is to permit a few Black women to acquire positions of authority in institutions that legitimize knowledge and to encourage them to work within the taken-for-granted assumptions of Black female inferiority shared by the scholarly community and the culture at large.

\cite{collins1989social}
\end{quote}
\vspace{1em}

Many Black women, as many people generally, arrive at the computational sciences without much formal training in detecting and tackling systemic oppression. 
Once inside the system, they are pressured to acquiesce to the status quo and cultivate ignorance or at least tolerance of systemic oppression. 
Black women are rewarded for capitulating to racist and misogynist norms, while also getting punished, often subtly, for minor dissent or missteps \citep{collins1989social}.
These select few Black women are tokenised by the self-preservation mechanisms of the system. 
They are allowed access to positions of power, although often merely impotent ceremonial roles, in order to appease those who request equity, diversity, and inclusivity.
``Those Black women who accept [the system] are likely to be rewarded by their institutions [but] at significant personal cost.'' This does not mean that Black women are passive recipients of systemic injustice. Far from it, many actively oppose and push back against it. Nevertheless, ``those challenging the [system] run the risk of being ostracized.'' \citep[p. 753][]{collins1989social}

The structural and interpersonal components of computational sciences make it difficult (if not impossible) for Black women to describe (let alone navigate, survive, or flourish in) their environment.
This results in confusion, and abuse, and confusion about abuse: a form of systemic-level gas-lighting. 
Ultimately, it can also lead to Black people making a Faustian pact in order to ensure their individual survival within this ecosystem: trade any preexisting principles they have --- or adopt the white man's principle as their own \cite[]{freire1970pedagogy} as the academic ecology trains them not to know any better --- thus, aligning them with male and white supremacy.
This results in the almost bizarre case wherein the few, highly tokenised (both with and without their consent and realization), Black women are not in any way directly contributing to the dismantling of the forces which keep their fellow Black women excluded \citep{collins1989social}.
In other words, if not trained in critical race studies and other critical fields, a Black computational scientist risks producing the same oppressive, hegemonically-aligned work, as any other, e.g., white, scientist.
Black women face a challenge, a dilemma, between: $a$) telling their truth (i.e., challenging the orthodoxy) and facing silencing, exclusion, and censorship at the institution and system levels (i.e., through the marginalisation of their work); or $b$) working to maintain the status quo which overtly rewards them yet covertly coerces them into supporting a system that devalues their humanity \citep[][]{OkunND}.

Privileged people are left unscathed by the nuanced and system-level issues we touch on herein.
Furthermore, these issues are difficult to acknowledge for those in power --- they are seen as a sideshow, a political/politicised distraction rather than an essential element of good (computational) science.
Alas, even when acknowledged the common mitigation is the creation of so-called diversity boards, which are often composed predominantly of white women.
And as we discuss above, white women can be part of the problem, especially when they enact white feminism.
This results in (further) tokenization of Black women and other minoritised groups.
Compounding these issues even further, although the active inclusion of Black women can be part of the solution, we argue that it can also be problematic, even leading to further exacerbating problems.
For two reasons: $a$) it gives the illusion that the inclusion of individuals can alone solve structural and deep-rooted problems; and $b$) the selected individuals themselves, although from a minoritized group, might not be equipped to recognize and tackle systemic oppression due to their academic training, harming both themselves and other minoritized groups that they are supposed to represent and help.
In other words,  we oppose the prevalent individual-centred solutions to systemic problems.
In considering the lack of Black women, a shift is required in the core questions we ask ourselves --- from the misguided ``why are Black women not entering computational sciences?'' to questions like ``what should the field as a whole, and computational departments specifically, do to create a welcoming and nurturing environment for Black women?''

The active inclusion and respectful representation of Black women is key to their safe progression in academia.
We all need to ``recognize the scale and scope of anti-Blackness'' within the computational sciences \citep{guillory2020combating}. 
However, promoting representation and/or inclusion, without acknowledgement of how white supremacy works and without challenging structural inequalities, is doomed to fail.
And as we saw, Black people themselves could be victims, unable to see outside their conditioning, and predominantly thinking in a manner that benefits white supremacy.
A representative demographic make-up should be seen more as the side-effect, the byproduct, of a healthy system and not an ingredient by which to bring such a system about.
Visible representation matters, but only if the ecosystem is set up to welcome and retain minoritised groups without exploiting them \citep[][]{Berenstain2016, Sloane2020}.



\section{Conclusion}

\begin{quote}
Freedom is acquired by conquest, not by gift. 

\cite{freire1970pedagogy}
\end{quote}
\vspace{1em}




Individual level issues such as interpersonal displays of racism are not the cause but a side-effect, symptomatic of a much deeper problem: structural, systemic, social, and institutional racism and sexism --- ideals and values set in place purposefully a couple of centuries ago \citep{Saini2019}.
Individual acts would be punished, or least outlined as things better avoided, if the current academic system was aligned with decolonisation instead of white supremacy.
Indeed part of the longevity of the system of promoting whiteness and masculinity to the detriment of Black women is exactly this: only those who support masculine and white hegemony ``float'' to the top.
Any members of minoritised groups, e.g., Black women, are often specifically selected (through systemic forces) to be trainable into upholders of the status quo --- conditioned to uphold currently extant kyriarchal \citep{Schussler2009} structures.
Those Black women who ``make it'' without buckling under pressure, face 
interpersonal and systemic abuse. 
And any work they do contribute to, any scientific progress they lead or take part in, is also systemically erased, forgotten --- disallowing them in large part from even becoming role models for others, for example, see \autoref{fig:mouton} \citep{Nelsen2017}.

The continuity of history is apparent both in terms of current research themes as well as in terms of present-day field-wide demographics.
Present-day academic oppression is often nuanced, covert, even imperceptible to most, including minoritized groups.
To some extent, we are all products of an academic tradition that trains us to conform to the status quo, almost by definition.
Continued critical engagement and enrichment of our vocabularies are necessary to articulate our oppressions and experiences, allowing us to overcome conditioned and internalized white supremacy, racism, and coloniality.  
Reevaluating our understanding of our fields' histories is paramount --- both the good \citep[e.g., Black women such as Melba Roy Mouton, see \autoref{fig:mouton} and][]{Nelsen2017, bwic_2016} and the bad \citep[e.g., eugenics and race science, expelling women from computational sciences and the tech industry, etc.,][]{Saini2019, Hicks2017}.

Academia produces work that predominantly maintains the status quo.
Those who push back against this orthodoxy are met with hostility, both at systemic and individual levels.
Majoritarian and minoritised people alike, who conform to the core values of racism, colonialism, and white supremacy are rewarded.
The promotion of people who are ideologically aligned with the current hegemony is how the system sustains itself --- both directly through the tenure system and generally through who is allowed into science and which roles and opportunities are open to them \citep[][]{Gewin2020}.

Ultimately, decolonising a system needs to go hand-in-hand with decolonising oneself. Structural obstacles (through the form of racism, coloniality, white supremacy, and so on) which prevent Black women and other minoritized groups from entering (and remaining in) computational sciences need to be removed.
At minimum, this requires the beneficiaries of the current systems to acknowledge their privilege and actively challenge the system that benefits them.
This is not to be confused with asking those in positions of power to be generous or polite to Black women nor are Black women passively asking for a ``handout'' or special treatment.
The healthy progression of computational sciences is one that necessarily examines, learns from, and dismantles its historical and current racist, colonialist, and oppressive roots, albeit through a gradual process.
Such a journey is beneficial not only to Black women but also to science in general.
Nonetheless, it is paramount to acknowledge the present ecosystem of the computational sciences for what it is and obtain our liberation from our conditioned internalized coloniality, white supremacy, and Anglo- and Euro-centrism.
These demands need to necessarily emerge from within.
``The liberation of the oppressed is a liberation of women and men, not things. Accordingly, while no one liberates [themselves] by [their] own efforts alone, neither [are they] liberated by others.'' \citep{freire1970pedagogy}



\section{Acknowledgement}
Abeba Birhane was supported, in part, by Science Foundation Ireland grant 13/RC/2094 and co-funded under the European Regional Development Fund through the Southern \& Eastern Regional Operational Programme to Lero, the Science Foundation Ireland Research Centre for Software, Ireland (www.lero.ie)

Olivia Guest was supported by the Research Centre on Interactive Media, Smart Systems and Emerging Technologies (RISE) under the European Union's Horizon 2020 programme (grant 739578) and the Republic of Cyprus through the Directorate General for European Programmes, Coordination and Development.

The authors would like to thank
Pinar Barlas,
Sebastian Bobadilla Suarez,
Johnathan Flowers,
Timnit Gebru,
Mustafa I. Hussain,
Saif Asif Khan,
Saloni Krishnan,
Andrea E. Martin,
Stephen Molldrem,
Alexandre Pujol,
Elayne Ruane, 
Iris van Rooij,
Luke Stark,
Reubs Walsh,
and others for discussing and/or commenting on earlier versions of this manuscript.

A version of this work will appear in the Danish Journal of Women, Gender and Research (https://koensforskning.soc.ku.dk/english/kkof/) in December 2020.

\bibliography{bibliography}

\end{document}